\documentclass{sigchi-ext}
\usepackage[T1]{fontenc}
\usepackage{textcomp}
\usepackage[scaled=.92]{helvet} % for proper fonts
\usepackage{graphicx} % for EPS use the graphics package instead
\usepackage{balance}  % for useful for balancing the last columns
\usepackage{booktabs} % for pretty table rules
\usepackage{ccicons}  % for Creative Commons citation icons
\usepackage{ragged2e} % for tighter hyphenation

\def\plaintitle{SportsXR - Immersive Analytics in Sports} 
\def\emptyauthor{}
\def\plainkeywords{Immersive analytics; sport analytics; visual analytics; information visualization.}

\title{SportsXR - Immersive Analytics in Sports}

\numberofauthors{4}
\author{%
  \alignauthor{%
    \textbf{Tica Lin}\\
    \affaddr{Harvard University} \\
    \affaddr{Cambridge, MA 02138, USA} \\
    \email{mlin@g.harvard.edu} 
  }
  \alignauthor{%
    \textbf{Yalong Yang}\\
    \affaddr{Harvard University} \\
    \affaddr{Cambridge, MA 02138, USA} \\
    \email{yalongyang@g.harvard.edu} 
  }
  \vfil
  \alignauthor{%
    \textbf{Johanna Beyer}\\
    \affaddr{Harvard University} \\
    \affaddr{Cambridge, MA 02138, USA} \\
    \email{jbeyer@g.harvard.edu} 
  } 
  \alignauthor{%
    \textbf{Hanspeter Pfister}\\
    \affaddr{Harvard University} \\
    \affaddr{Cambridge, MA 02138, USA} \\
    \email{pfister@g.harvard.edu} 
  }
}

\definecolor{linkColor}{RGB}{6,125,233}
\hypersetup{%
  pdftitle={\plaintitle},
  pdfauthor={\emptyauthor},
  pdfkeywords={\plainkeywords},
  bookmarksnumbered,
  pdfstartview={FitH},
  colorlinks,
  citecolor=black,
  filecolor=black,
  linkcolor=black,
  urlcolor=linkColor,
  breaklinks=true,
}

\usepackage{url}

\usepackage{breakurl}
\begin{document}

%% For the camera ready, use the commands provided by the ACM in the Permission Release Form.
\CopyrightYear{2020}
\setcopyright{rightsretained}
\conferenceinfo{CHI'20,}{April  25--30, 2020, Honolulu, HI, USA}
\isbn{978-1-4503-6819-3/20/04}
\doi{https://doi.org/10.1145/3334480.XXXXXXX}
%% Then override the default copyright message with the \acmcopyright command.
\copyrightinfo{\acmcopyright}

\maketitle

% Uncomment to disable hyphenation (not recommended)
% https://twitter.com/anjirokhan/status/546046683331973120
\RaggedRight{} 

% Do not change the page size or page settings.
% !TEX root = ../SportsXR.tex

\begin{abstract}
%   We present our initial investigation of key challenges and potentials of immersive analytics (IA) in sports, or SportsXR. Sports have highly collaborative and strategic nature, making it well suited for immersive analytics. SportsXR can be used to eliminate the gap between domain experts and analytic experts, which has been the key issue most sport practitioners face now. We discuss challenges faced by IA research community, including sports data collection, analytics expertise, dynamic situated data visualization and interaction design, and domain experts collaboration. We then present potential areas for SportsXR application in training, coaching and fans experiences. The discussion aims to serve as an inspiration for future SportsXR research.

 We present our initial investigation of key challenges and potentials of immersive analytics (IA) in sports, which we call SportsXR. Sports are usually highly dynamic and collaborative by nature, which makes real-time decision making ubiquitous. However, there is limited support for athletes and coaches to make informed and clear-sighted decisions in real-time. SportsXR aims to support situational awareness for better and more agile decision making in sports. In this paper, we identify key challenges in SportsXR, including data collection, in-game decision making, situated sport-specific visualization design, and collaborating with domain experts. We then present potential user scenarios in training, coaching, and fan experiences. This position paper aims to inform and inspire future SportsXR research.
\end{abstract}


% ACM Classfication



% Print the classficiation codes
% Please use the 2012 Classifiers and see this link to embed them in the text: \url{https://dl.acm.org/ccs/ccs_flat.cfm}

% !TEX root = ../SportsXR.tex

\vspace{-0.5em}
\section{Introduction}
Immersive Analytics (IA) builds upon novel immersive technologies and extends data visualization and analytics capabilities beyond traditional desktop workspaces. 
 %IA provides advantages in exploration of multiple datasets with large display spaces, comprehending complex data through multi-sensory interface, collaboration across multiple users, and situated analytics right in the context of the events to eliminate the gap between people, data and tools~\cite{t_chandler_immersive_2015}. 
 %The power of IA stems from its inherent support for
For example, IA can facilitate the exploration of heterogeneous sports datasets by making use of large display spaces, providing multi-sensory interfaces, promoting expert collaboration, and enabling situated analytics. 
Situating the visualization right in the context of the data and events eliminates the gap between people, data, and tools~\cite{t_chandler_immersive_2015}. 
Potential applications of immersive analytics cover a wide range of domains, such as life and health sciences, construction site management, supply chain, and factory planning~\cite{marriott_immersive_2018}. 
Immersive analytics in sports has received relatively low attention so far.
 %However, there has been very little discussion around immersive analytics in sports. 
However, due to its highly collaborative and strategic nature, sports have a huge potential for immersive analytics.

\begin{marginfigure}
	\begin{minipage}{\marginparwidth}
	  \centering
	  \includegraphics[width=\marginparwidth]{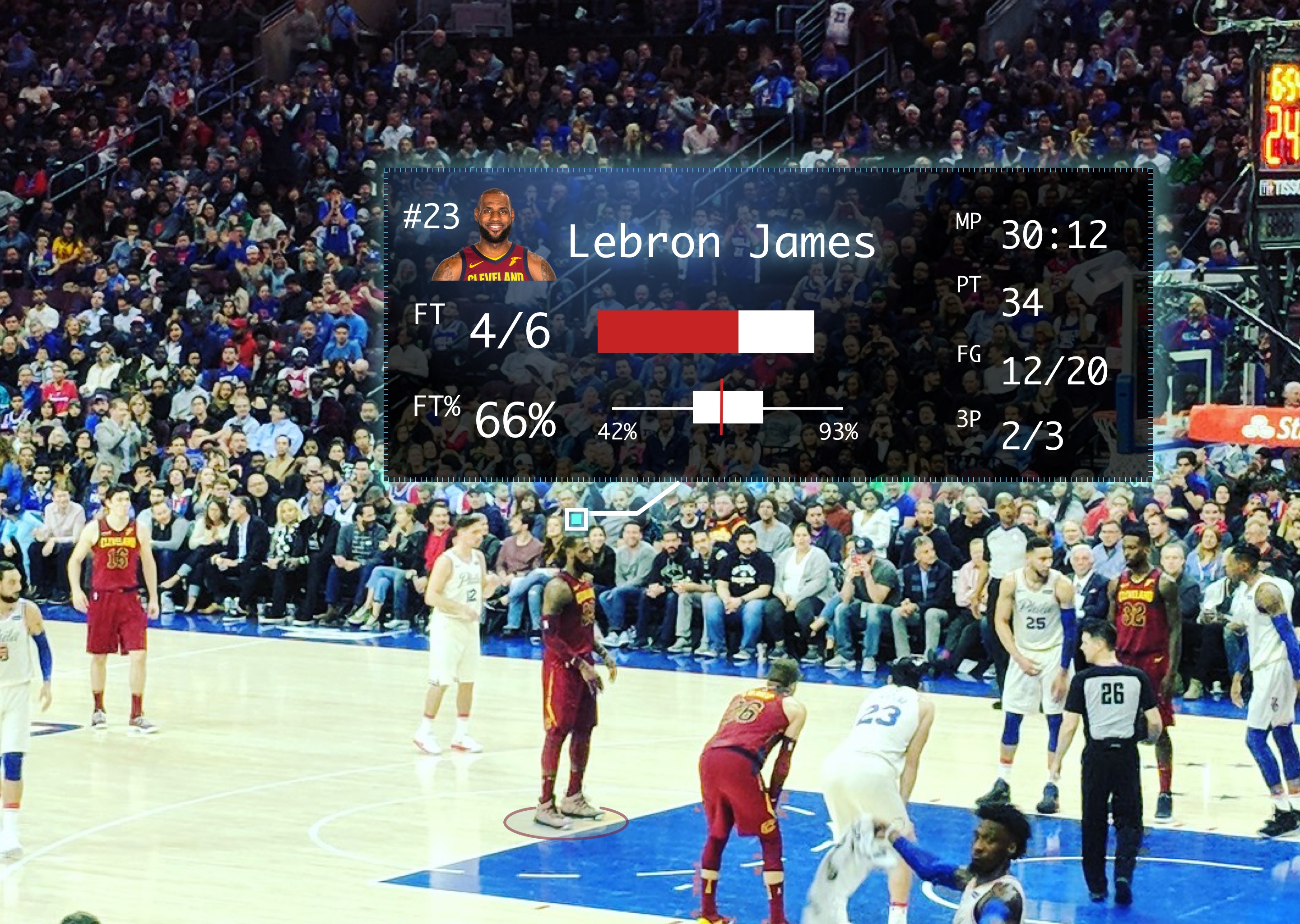}
	  \caption{A simulated basketball fan experience with player statistics shown in AR.}
	  \label{fan}
	\end{minipage}
\end{marginfigure}

Sports generate a huge amount of heterogeneous data, such as real-time positions of players, physiological measurements of players, tactical trajectories, and scouting insights. 
Current sports data visualization research mostly focuses on visual analytics systems on traditional 2D screens~\cite{Perin2018-jh}, which are usually used off-line. 
However, in practice, there is a huge demand for in-game decision making.  
Thus, a more situated and user-centered approach to visualize and analyze sports data is required to bring real-time analytics to decision-makers (e.g., players, coaches, and team executives). 
Applying IA to sports can engage a broader audience with embodied analytic capabilities and enable data-driven decision making in various use cases. For example, a real-time overlay of a player's box score and an intuitive data analysis interface on a heads-up display could bring fan engagement to the next level (see Fig.~\ref{fan}).

\begin{marginfigure}
  \begin{minipage}{\marginparwidth}
    \centering
    \includegraphics[width=\marginparwidth]{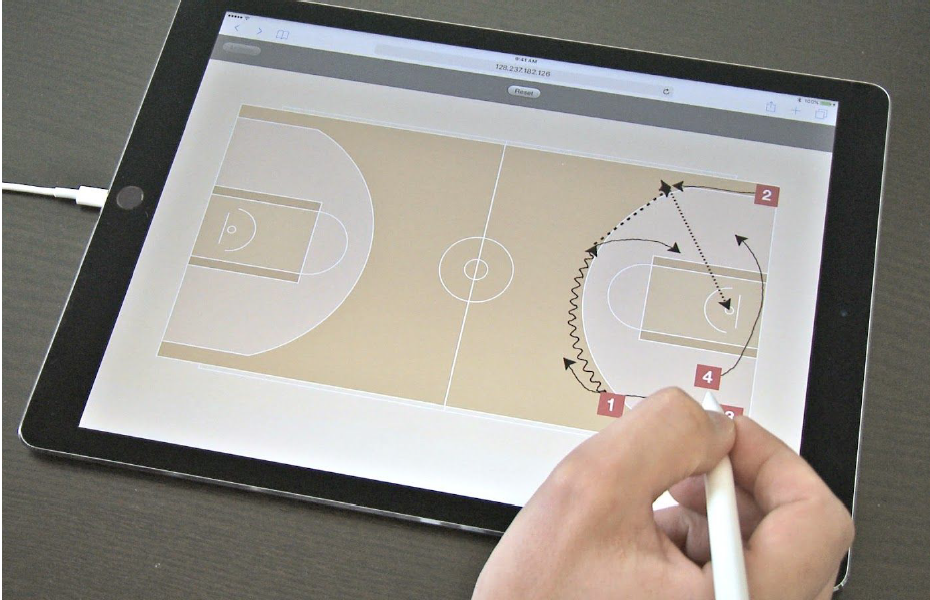}
    \caption{Interactive play sketching with synthesized defenses by Seidl et al.~\protect\cite{Seidl2018-ky}.}
    \label{seidl-2018}
  \end{minipage}
\end{marginfigure}

% Visualizations have been widely used to facilitate the analysis of sports data. Many visual analytic systems have been developed on traditional 2D screens~\cite{Perin2018-jh}.
% However, in certain applications, 2D screens reach their limit in terms of what they can represent (e.g. 3D shooting trajectories is difficult to analyze on 2D screens) and where they can be used (e.g. coaches may need interactive analytical reasoning embedded in a physical court). 
% SportsXR aims to use new display and interaction technologies to enable immersive and situated analytics in sports. 

In this paper, we present the challenges and potentials of applying IA to sports to eliminate the gap between sports data, people, and tools from the perspectives of IA researchers. 
We, furthermore, outline a vision of SportsXR in three specific user scenarios that empower athletes, coaches, and fans. % with analytic capabilities through spatial immersion, situated analytics, and collaboration.
% In this paper, we are going to discuss:
% \begin{itemize}
%     \item the current state of related research;
%     \item the challenges in SportsXR;
%     \item potential application scenarios of SportsXR.
% \end{itemize}

% !TEX root = ../SportsXR.tex

\section{Related Research Fields}
SportsXR builds upon a variety of research fields. At its heart, SportsXR integrates sports analytics with situated analytics. Situated analytics is a subset of the larger field of immersive analytics, which focuses on applying embodied display and interaction techniques (e.g., augmented and virtual reality) for data visualization and analytics. Situated analytics, in addition, emphasizes the use of augmented reality to link abstract visual representations to objects in the physical world.

\subsection{Sports Analytics}
Sports analytics has become popular with the success of the MLB Moneyball story in 2002. Oakland Athletics General Manager Billy Beane, with a limited budget, used sabermetrics to draft undervalued players and subsequently won games against rich teams like the New York Yankees~\cite{Lewis2004-zh}. 
The recent explosion of computational power, motion capture technology, and statistical methods make advanced sports analytics feasible in many scenarios. This includes game outcome prediction, measurement and evaluation of player performance, analysis of rules and adjudication, and within-game strategy. 
% through building predictive models on top of large data collection. 

% Some examples of using statistical models for strategies in sports include game outcome models, measurement and evaluation of player performance, analysis of rules and adjudication, and within-game strategy~\cite{American_Statistical_Association_issuing_body2005-bs}.

\begin{marginfigure}
  \begin{minipage}{\marginparwidth}
    \centering
    \includegraphics[width=\marginparwidth]{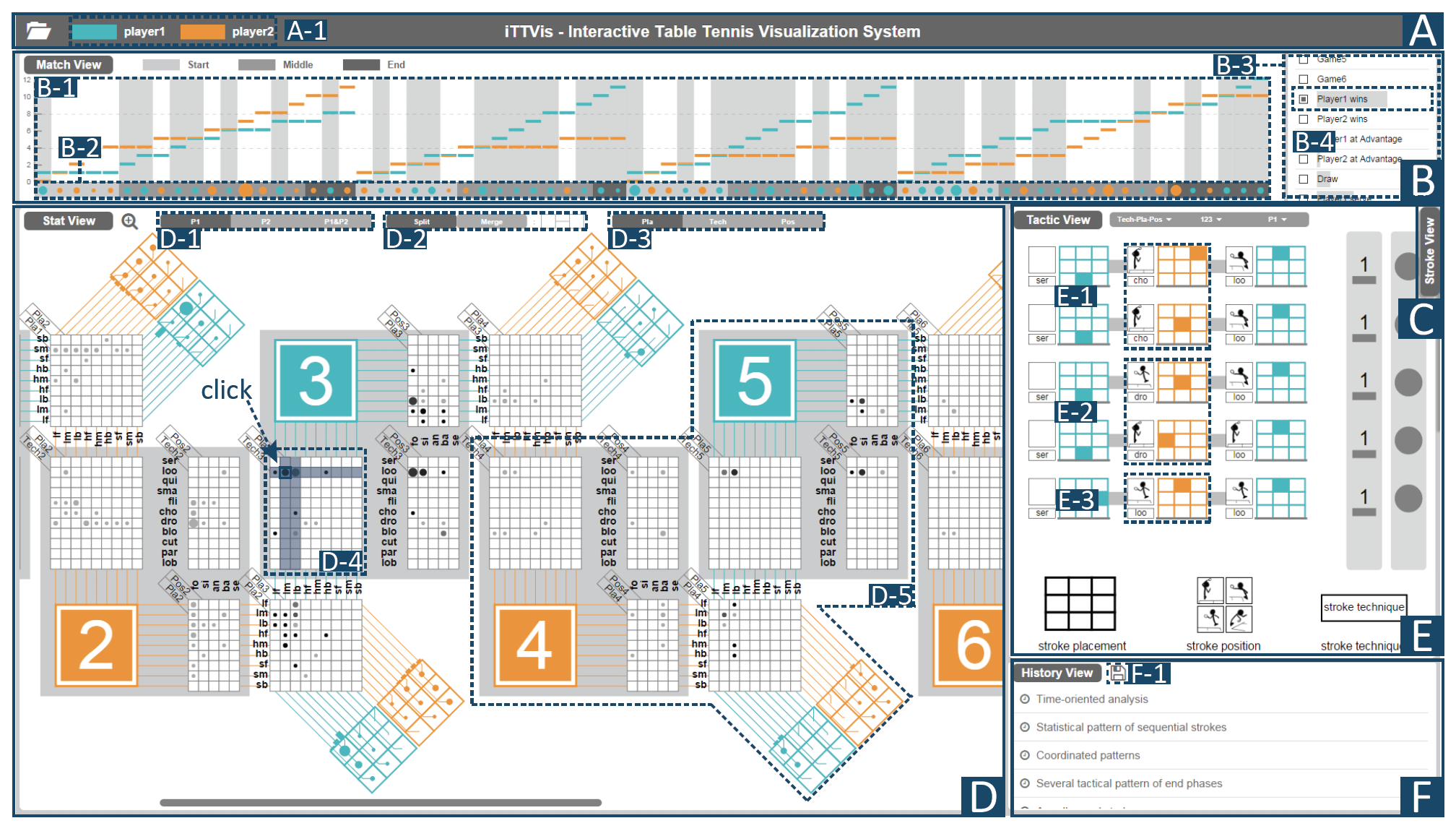}
    \caption{Interactive visualization of table tennis data by Wu et al.~\protect\cite{Wu2018-gf}. \textit{$\copyright$ 2018 IEEE, reprinted with permission}.}
    \label{wu-2018}
  \end{minipage}
\end{marginfigure}

However, these fully automatic methods do not integrate the knowledge of decision-makers (i.e., coaches, scouts, and managers) in an effective way~\cite{Alamar2013-ag}.
% analytics itself won’t change the game unless the decision makers  integrate the analytic insights with their knowledge in sports. 
% Much research interest of sports analytics turn toward sports 
Data visualization and visual analytics have been integrated into the sports analytics workflow to enable human-in-the-loop analytics. For example,
Seidl et al.~\cite{Seidl2018-ky} use NBA game tracking data to perform data-driven ghosting defense prediction. They also offer a tablet-based interface for coaches to perform real-time strategy planning (see Fig.~\ref{seidl-2018}). Wu et al.~\cite{Wu2018-gf} create a holistic visualization system for table tennis match data and empower domain experts to find unnoticed patterns (see Fig.~\ref{wu-2018}). 
Perin et al.~\cite{Perin2018-jh} have compiled a comprehensive survey of these systems. However, there is still a clear demand for more natural and transparent interface design to intuitively interact with sports data to perform meaningful data analytics and prediction.

% variety of sports data visualization articles both from industries and academia also indicate their importance in the sports analytics domain.

\subsection{Immersive and Situated Analytics}
\begin{marginfigure}
  \begin{minipage}{\marginparwidth}
  \centering
  \includegraphics[width=\marginparwidth]{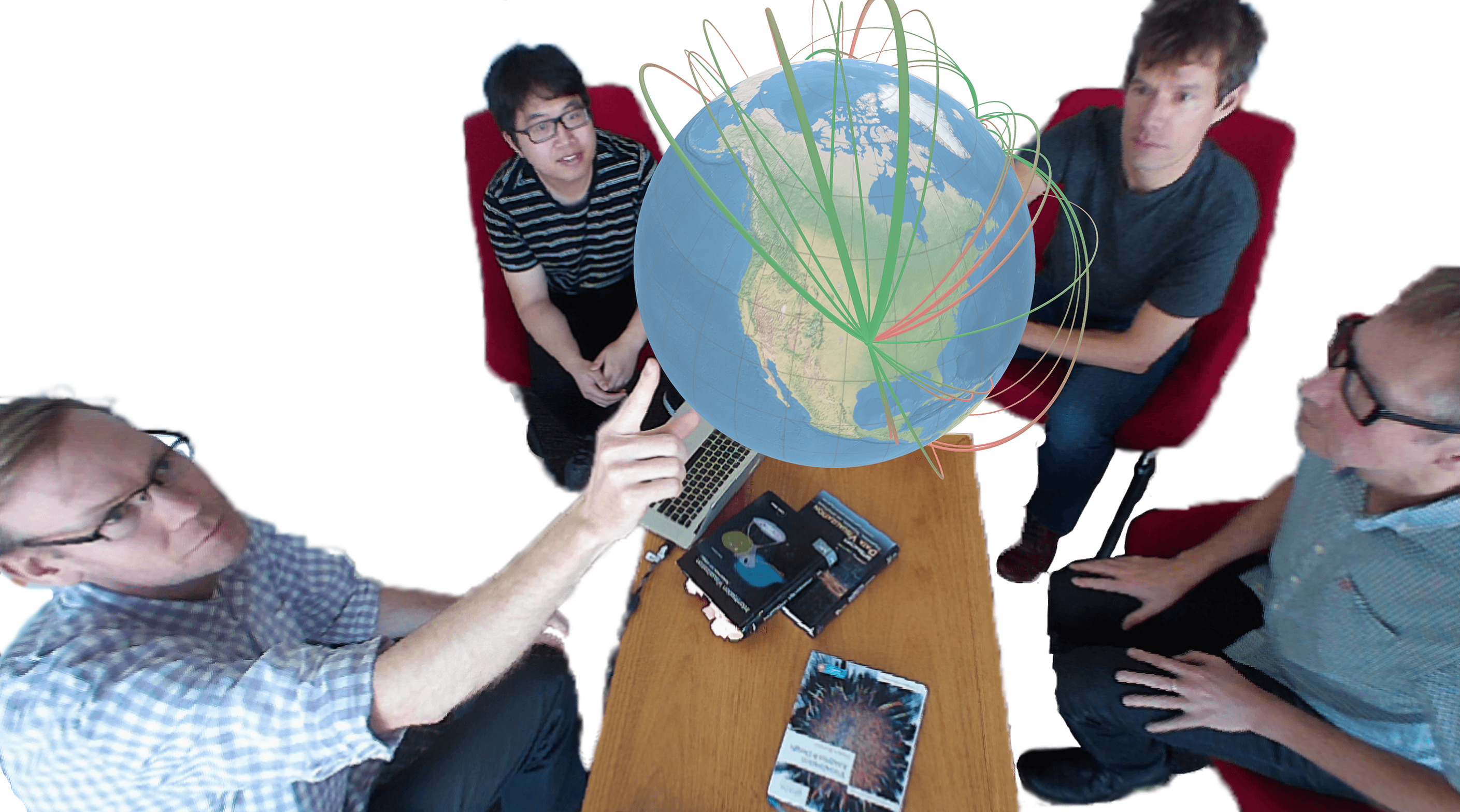}
  \caption{An in-situ collaborative scenario demonstrated by Yang et al.~\protect\cite{yalong_2019}. \textit{$\copyright$ 2019 IEEE, reprinted with permission.}}
  \label{yalong_2019}
  \end{minipage}
\end{marginfigure}

% Immersive analytics builds upon the development of immersive technologies and brings data visualization and analytics capabilities beyond traditional desktop work-spaces.
% \hl{shorten the description, more situated examples}
Immersive analytics leverage new display and interaction techniques, such as AR/VR headsets, for visual analytics~\cite{marriott_immersive_2018}.
%Compared to traditional displays, AR/VR headsets can render very large displays at low cost with a high degree of portability (compared to wall displays and CAVEs), thereby creating immersion through situated and collaborative visualization. 
Compared to traditional displays, AR/VR headsets can render very large displays at low cost with a high degree of portability compared to wall displays and CAVEs. Therefore, AR/VR headsets can create immersion through situated and collaborative visualization.
This opens up opportunities for direct interaction with visuals that are radically different from desktop computing paradigms. 
% Immersive analytics leverage new display and interaction techniques for visual analytics~\cite{marriott_immersive_2018}.
% Virtual and Augmented Reality (VR and AR) headsets are commonly recognized as the representative devices for Immersive Analytics. 
% Compare to traditional displays, AR/VR headsets can render (1) very large displays at low cost with (2) a high degree of portability (compared to wall displays and CAVEs) and (3) immersiveness for (4) situated and (5) collaborative visualisation, and (6) new kinds of direct interaction with visuals that are radically different to desktop computing paradigms. 
% Many immersive analytics research have been conducted in recent years. 
Situated analytics is a form of in-situ interactive visual analysis. In situated analytics, visual representation of information is immediately linked to physical objects to facilitate sense-making~\cite{marriott_situated_2018} (e.g., see Fig.~\ref{yalong_2019}). 
% Physical objects often associate important information for sense-making. 
% Situated analytics provide visual representation of information immediately linked to physical objects to help people perceive information better. 
% For example, \hl{describe and cite a few papers with use cases}.

% Immersive analytics provide advantages in exploration of multiple datasets with large display spaces, comprehending complex data through multi-sensory interface, collaboration across multiple users, and situated analytics right in the context of the events to eliminate the gap between people, data and tools. The discussion around the application of immersive analytics has covered a wide range of potential domains, such as life and health sciences, construction sites management, supply chain and factory planning, economic research, etc. However, there has been very little discussion around immersive analytics in sports, yet due to highly collaborative and strategic nature, sports preserve large potentials for immersive analytics.

% \subsection{VR in Sports}
VR for sports training has been preliminarily explored, e.g., for basketball and baseball:
Tsai et al.~\cite{Tsai2019-ns} use a VR simulation to train basketball players' decision-making skills, and Zou et al.~\cite{Zou2019-vq} visualize simulated pitch and bat swing data in VR to improve a baseball player's batting eye. 
%VR devices were used in these cases for their immersion and situated capabilities. 
Applying AR in sports is largely unexplored and can bring unique opportunities for the integration of digital information and physical movements.    

% 

% The gap between data, people and tools exists in sports due to the diverse and complex nature of sports data, highly context-driven, situated and human-in-the-loop decision making, and requirement of close collaboration between coaches, analysts and athletes. The high demands on data understanding, situated data analytics and collaborative data-driven decision making make sports a great candidate for immersive analytics. 

% In this paper, we present the challenges and potentials of applying immersive analytics to sports to eliminate such gap from the perspectives of IA researchers. We end the discussion by demonstrating the vision of SportsXR to empower domain experts with analytic power through spatial immersion, situated analytics and collaboration in immersive analytics~\cite{t_chandler_immersive_2015}.

%%% Newly added 
\subsection{Industrial Sports Analytic Platforms}
There are several innovative commercial products applying AR/VR for sports training, coaching, and fan experiences. STRIVR ~\cite{strivr} uses VR as an immersive learning platform to simulate real game scenarios to train decision making and mental preparation for professional football players, skiers, and basketball players. Rezzil~\cite{rezzil} provides similar VR training with coaching and analytic features for professional soccer players. Second Spectrum~\cite{second_spectrum} experiments on player tracking and AR features for visual information augmentation on broadcasting and video re-play. CourtVision~\cite{courtvision} provides information and stats overlays in AR for enhancing live sports viewing experiences for NBA basketball fans within minutes of the live game. 
% However, gaps in providing analytic supports in real-time still exist.
However, those applications only cover some sub-areas of SportsXR, leaving many other areas still being unexplored.
%. There are many other unexplored applications. 

% !TEX root = ../SportsXR.tex

\section{Challenges}
We have identified four main challenges in SportsXR research: 1) sports data collection; 2) in-game decision making; 3) situated sports-specific visualization design; and 4) collaboration with domain experts.

% The challenges we identify to perform successful SportsXR research include sports data collection, analytic expertise, dynamic situated data visualization and interaction design, and domain experts collaboration.

\subsection{Sports Data Collection and Extraction}
Sports data are inherently complex. Typical datasets contain a combination of heterogeneous, multi-dimensional, and unstructured data. Examples include box score data, tracking data, scouting reports, game video clips, players' mentality, and many other qualitative metrics~\cite{Perin2018-jh}. 
%Sports data are inherently complex, containing not only well-structured data like box score data, tracking data, and meta-data~\cite{Perin2018-jh}, but also heterogeneous data, such as scouting reports, game video clips, players' mentality, and many other qualitative metrics.
% human decision making~\hl{not sure do you think calling it structured data and un-structured data makes more sense?}, collaboration and individual physical differences over time,
% which makes data collection, visualization and analytics difficult to process. 
%Furthermore, it is difficult to ensure data collection is satisfactory and unbiased for the desired analytic work flow. 
Standards for data collection need to be enforced to ensure that the collected data is unbiased and compatible with the subsequent analytic workflow. 
Meeting these standards requires considerable efforts for real-time data recording, pre-processing, data cleaning, and formatting. 
Furthermore, in the context of situated and in-game analytics, data often needs to be extracted from video feeds and live sensors. This, in turn, requires state-of-the-art computer vision techniques to track game objects and players.
%
%and applying computer vision techniques to extract information.
% of sports data that are satisfactory for accurate sports analytics 
%differ by sport types and by context, and data collection also requires considerable preparation efforts for real-time data recording, pre-processing, cleaning, formatting and applying computer vision techniques to extract information. 
For example, basketball analytics collects dynamic movement data of all players and detects shot types and defensive vs. offensive tactics to evaluate the performance of each player, while baseball analytics focuses more on pitch type and base rate to evaluate player value. 
% \hl{maybe mention the need of computer vision here.}

The first challenge, therefore, lies in identifying, collecting, and extracting various data required for different sports and contexts. 

\subsection{In-game Decision Making}
% Ever since the Oakland Athletics baseball team proved the value of modern sports analytics in 2002, there has been a continuous trend in major professional sports to develop own in-house analytic teams to support team-building and game strategy. 
Research on optimal statistical models and visualization tools for game prediction and evaluation in various sports has seen a huge growth in the last decade. However, current sports analytics only occur asynchronously in the hands of data scientists, leading to inefficient communication and inaccurate decision making on the athlete side. By bringing in-situ analytic power to domain experts such as coaches and athletes, immersive analytics will play an important role in eliminating the gap between data-driven strategies and executions. 

The second challenge, therefore, lies in making analytics models not only understandable to coaches and athletes but to also enable fast in-game and in-situ analytics. %and available to coaches and athletes in-game. 
%This also brings up the needs of close collaboration with domain experts for the design of sports-specific visualizations.

\subsection{Situated Sport-specific Visualization Design}
% With successful data collection, analytic models and accurate insights of analytic requirements in sport workflow, the consequent design of immersive data visualization and interaction also presents its unique challenges. 
%While some immersive analytics can be performed regardless of the actual realities, SportsXR requires the interaction and analytics to be performed in-situ and be updated dynamically to maximize its impact.
One distinguishing feature of SportsXR is that analytics need to be performed in-situ and need to be updated dynamically to maximize its impact.
However, coaches and athletes primarily have to focus on real-world movements and events.
%the attention of a coach or athlete will mostly focus on real-world movements and situations. 
Thus, the visualizations in SportsXR need to be situated, highly dynamic, concise, and context-dependent to play an auxiliary role. For example, a basketball coach using SportsXR to evaluate offensive options needs different analytics based on a player's location and a defender's movement. A spatial shot percentage map of the ball handler might be helpful when the defense is wide open, but when the player is being contested, the shot percentages of other shooters on the team are more important. 
Deciding when and how to change the presented data to optimally support data-driven decision making will be a key design consideration in SportsXR. 

The third challenge, therefore, lies in defining the context-dependent factors based on real-world events to decide how to present and interact with dynamic situated data.

\subsection{Collaboration with Domain Experts}
% \hl{Maybe title as ?}
% Similar to other visualization and immersive analytics research, the close collaboration with domain experts stays at the top for a successful problem-driven research. 
User-centered design has been proven to be an effective way of engaging end-users~\cite{lam_bridging_2018}.  
Batch et al.~\cite{Batch2019-wc} worked closely with economic data analysts to integrate immersive visualization into their actual workflow. 
In contrast, domain experts in sports rarely perform data analytics themselves but make in-game decisions based on experience rather than based on data. 
%However, the same collaboration process presents more difficulties from that the domain experts in sports rarely perform data analytics explicitly and make decisions based on experiences and insights rather than data. 
To understand the decision-making workflow of coaches and athletes, researchers first need to gain in-depth domain knowledge. Based on this domain knowledge, they can then identify and extract analytic components from the heuristic insights in order to improve the accuracy of their decision making. 
Furthermore, researchers need to educate their sports collaborators on analytics and visualization methods, as well as introduce them to novel AR interfaces for situated analytics.

In our current research project, we collaborate closely with both Harvard Men's and Women's Basketball teams to gather first-hand expert knowledge. Building personal sport expertise and working with data analysts on an interdisciplinary sports analytics team can lead to more effective collaborations with coaches and players. 

The fourth challenge, therefore, lies in establishing an effective collaboration between immersive analytics researchers and sports experts. 

\section{Potential User Scenarios}
% Innovative technologies are constantly applied to improve athletes performance and sports development, including the use of wearable and sensing technologies, computer vision, machine learning and data visualization. SportsXR can be used to eliminate the existing gap between data analytics and empirical insights, and to improve athletic and strategic performance through data-driven decision making. 
% \hl{will be better if we can link the potentials with the challenges somehow}
We have identified the following SportsXR scenarios: athlete training, coaching, and fan experience.

%propose the potential research areas of SportsXR in the following categories -- training, coaching and fans experience. 

\subsection{Training}
\begin{figure}
  \centering
  \includegraphics[width=\linewidth]{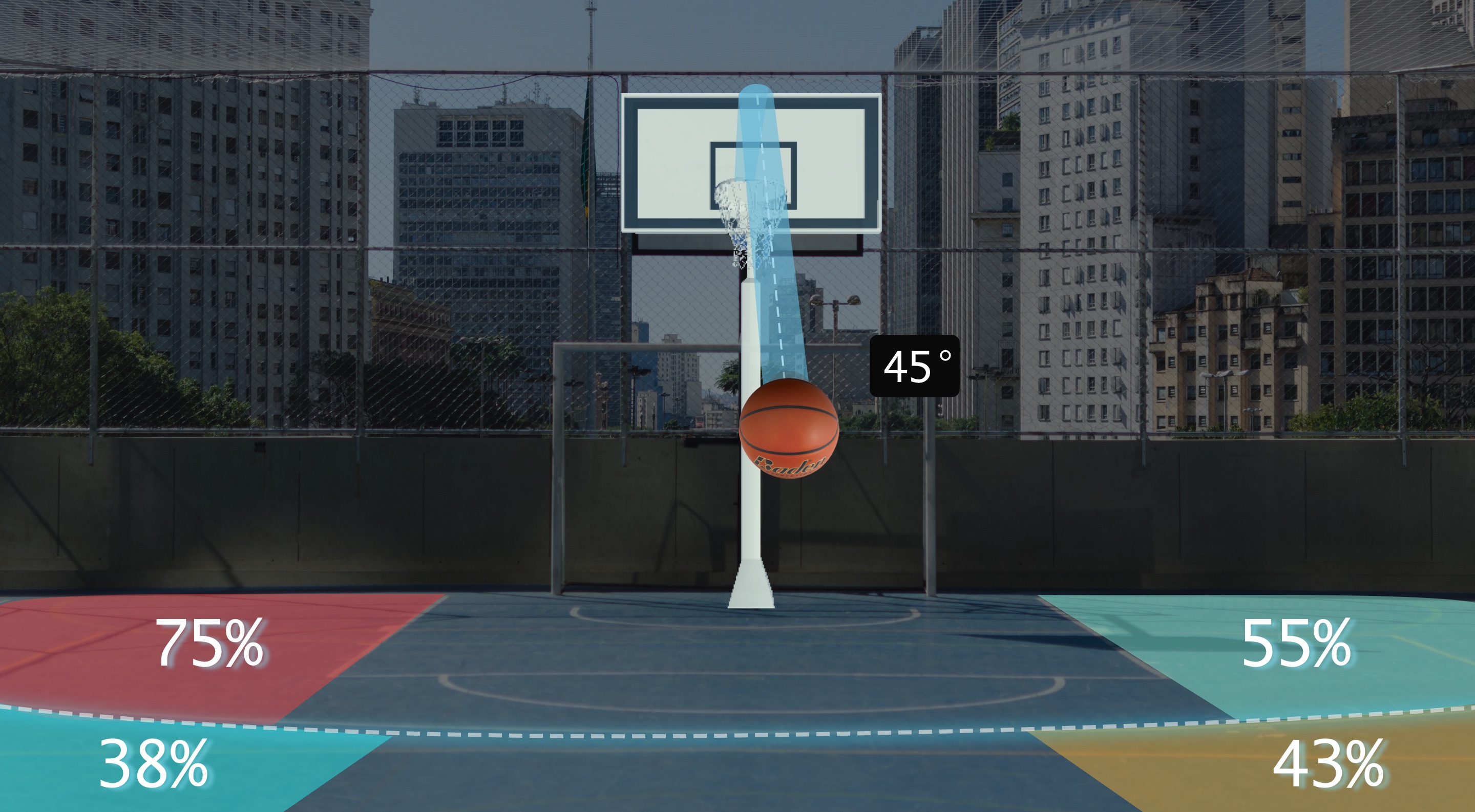}
  \caption{Basketball shooting training in SportsXR.}
  \label{training}
\end{figure}

Technological innovations in sports training are not just crucial to improve training effectiveness with instant performance feedback and overall body condition monitoring, but also to reduce injuries and to excel former records. 
Most data collected during training are interpreted and communicated to athletes by their coaches. However, embodied data analytics can empower athletes with the ability to self-evaluate and modify their techniques in real-time. Furthermore, immersive visualizations can enhance the collaboration between coaches and athletes by embedding the visuals spatially into the real world (see Fig.~\ref{training}). 
%
%While most data collected during training are interpreted and communicated to athletes by their coaches, immersive analytics has unique capabilities to empower athletes with embodied data analytics to allow self-evaluation and modification in real-time, and enhance the collaboration between coaches and athletes through immersive visualizations (see Fig.~\ref{training}). 
%
A pioneer example of this was demonstrated by Eliud Kipchoge breaking the 2-hour marathon barrier with the help of a projected pacing visualization in real-time~\cite{Eliud}.
%in the breaking of 2-hour marathon barrier by Eliud Kipchoge with the help of projected pacing visualization in real time. 

\subsection{Coaching}
So far, sports analytics has had its highest impact on coaching decision making. Data scientists in the backroom apply advanced machine learning and computer vision to convert video and statistical data into strategic insights. On top of the advanced statistical models that track detailed winning factors in players’ performance, the close collaboration between analysts and coaches has been the key factor for the successful team rebuild of the NBA Philadelphia 76ers~\cite{76ers}. 
However, for many teams and individual sports like tennis or track-and-field, employing a professional data analytics team is unfeasible. 
%For teams and sports with little professional analytic support, such as individual sports like tennis or track-and-field, it is unfeasible to achieve the same level of personnel structure. 
The spatial immersion and in-situ decision making of situated analytics allow SportsXR to embed analytics into the coaching workflow for a seamless integration of analytic and coaching insights.
%With situated analytics and collaboration through spatial immersion, SportsXR allows to embed analytics into the coaching workflow for a seamless integration of analytic and coaching insights.

\subsection{Fan Experience}
Immersive technologies have gradually become a big part of social experiences, such as AR filters on Snapchat or Instagram and VR game watching of the FIFA World Cup. To bring fan experience to the next level, personal storytelling through data visualization and content creation plays an important role in deepening a sports fan's engagement. SportsXR can add digital information overlays in real-time, and provide a more engaging interactive experience such as video annotation, dynamic data lookup, performance comparison, and customized view manipulation. The design challenges for immersive visualization and interaction, in this case, are largely unsolved research questions.

% !TEX root = ../SportsXR.tex

\section{Conclusions}
In this paper, we discuss the trends and challenges IA research communities are facing in the new research field of SportsXR. We have outlined potential research areas and encourage the exploration of SportsXR applications in training, coaching, and fan experience. Building upon the continuous research efforts in immersive analytics, \mbox{SportsXR} presents unique challenges for research: 
First, sports data collection is challenging due to complex data types and contexts. 
Second, analytics models need to be suitable for fast, situated decision making by athletes and coaches. % expertise challenge faced by domain experts, 
Third, SportsXR development requires the design of novel, sports-specific, dynamic, and situated data visualization and interaction methods, and fourth, SportsXR relies on a close collaboration between researchers and sports domain experts.
We hope that SportsXR will eliminate gaps between analytic and athletic insights, propel innovation in sports, and make sports analytics available to larger audiences.
%The potential of SportsXR to eliminate gaps between analytic and athletic insights can propel innovation in sports and empower sports analytics to benefit larger audiences.

\section{Acknowledgments}
We wish to thank Coach Kathy Delaney-Smith, Mike Roux, Mark Kaliris, and Lindsay Werner at Harvard Women's Basketball, and Mike Sotsky and Casey Brinn at Harvard Men's Basketball for their time and expertise. This research is supported in part by King Abdullah University of Science and Technology (KAUST) and the KAUST Office of Sponsored Research (OSR) award OSR-2015-CCF-2533-01.
\balance{} 

\bibliographystyle{SIGCHI-Reference-Format}
\bibliography{sportsXR-base}

\end{document}